\documentclass{article}

\PassOptionsToPackage{numbers, compress}{natbib}

\usepackage[final]{neurips_2025_ml4ps}

\usepackage[utf8]{inputenc}
\usepackage[T1]{fontenc}
\usepackage[dvipsnames]{xcolor}
\usepackage[colorlinks=true, allcolors=Blue]{hyperref} 
\usepackage{url}
\usepackage{booktabs}
\usepackage{amsfonts}
\usepackage{nicefrac}
\usepackage{microtype}

\usepackage{array}
\usepackage{xurl}
\usepackage{graphicx}
\usepackage{amsmath}
\usepackage{multirow}
\usepackage{amssymb}
\usepackage{mathtools}
\usepackage{amsthm}
\usepackage{doi}

\newcommand{\Sec}[1]{{\protect\hyperref[sec:#1]{Section~\ref*{sec:#1}}}}

\newcommand{\Fig}[1]{{\protect\hyperref[fig:#1]{Figure~\ref*{fig:#1}}}}

\newcommand{\subFig}[2]{{\protect\hyperref[fig:#1]{Figure~\ref*{fig:#1}~#2}}}
\newcommand{\Equ}[1]{{\protect\hyperref[equ:#1]{Equation~\ref*{equ:#1}}}}

\newcommand{\Tab}[1]{{\protect\hyperref[tab:#1]{Table~\ref*{tab:#1}}}}

\newcommand{\App}[1]{{\protect\hyperref[app:#1]{Appendix~\ref*{app:#1}}}}

\newcommand{\Alg}[1]{{\protect\hyperref[alg:#1]{Algorithm~\ref*{alg:#1}}}}

\title{Blind Strong Gravitational Lensing Inversion: Joint Inference of Source and Lens Mass with Score-Based Models}

\author{
  Gabriel Missael Barco$^{1, 2, 3}$ \quad Ronan Legin$^{1, 2, 3}$ \quad Connor Stone$^{1, 2, 3}$ \\ \textbf{Yashar Hezaveh}$^{1, 2, 3, 4, 5, 6}$ \quad
  \textbf{Laurence Perreault-Levasseur}$^{1, 2, 3, 4, 5, 6}$ \\
  $^1$Université de Montréal \quad $^2$Mila \quad $^3$Ciela Institute \quad $^4$CCA, Flatiron Institute \\
  $^5$Perimeter Institute for Theoretical Physics \quad $^6$Trottier Space Institute\\
  \texttt{\{gabriel.missael.barco,ronan.legin,connor.stone,yashar.hezaveh,}\\ \texttt{laurence.perreault.levasseur\}@umontreal.ca}
}

\begin{document}

\maketitle

\begin{abstract}
    Score-based models can serve as expressive, data-driven priors for scientific inverse problems. In strong gravitational lensing, they enable posterior inference of a background galaxy from its distorted, multiply-imaged observation. Previous work, however, assumes that the lens mass distribution (and thus the forward operator) is known. We relax this assumption by jointly inferring the source and a parametric lens-mass profile, using a sampler based on GibbsDDRM but operating in continuous time. The resulting reconstructions yield residuals consistent with the observational noise, and the marginal posteriors of the lens parameters recover true values without systematic bias. To our knowledge, this is the first successful demonstration of joint source-and-lens inference with a score-based prior.
\end{abstract}

\section{Introduction}

Score-based models have been successfully applied as data-driven, expressive priors for inverse problems. For example, in the field of astrophysics, they have been used for interferometric imaging \citep[e.g.][]{feng2023score, dia2025iris}, strong gravitational lensing source reconstruction \citep[e.g.][]{adam2022cla, Karchev2022}, cosmological-field inference \citep[e.g.][]{Legin2023, Ono2024,  Floss2024}, deconvolution \citep[e.g.][]{xue2023, Adam2025}, and many other applications.

These applications depend on approximations or heuristics, since posterior sampling with score-based priors is, in general, intractable.
Existing methods can be broadly classified into four categories \citep{zheng2025inversebench}: guidance-based methods \citep[e.g.][]{adam2022cla, song2023pseudoinverseguided, chung2023twee}, variable splitting \citep[e.g.][]{Wu2024}, variational Bayes \citep[e.g.][]{feng2023score, feng2024variational}, and sequential Monte Carlo \citep[e.g.][]{Dou2023}. In this work, we focus on the first category, in which an approximate likelihood term $\nabla_{\mathbf{x}_t}\log p_t(\mathbf{y} \mid \mathbf{x}_t)$ is used to guide the diffusion of the prior.

Most of these approaches assume that the parameters of the forward model are known, which is often not the case in practice. Jointly inferring the operator and the underlying parameters of interest (also known as \textit{blind inversion} in the literature \citep[e.g.][]{Levin2009blind, Gao2021blind_gem}) is an active area of research. Several approaches approximate blind-inversion sampling with diffusion models; for example, GibbsDDRM \citep{murata2023gibbs}, BlindDPS \citep{chung2023blinddps}, BIRD \citep{chihaoui2024bird}, and Fast Diffusion EM \citep{laroche2024fastem}.

Strong gravitational lensing, which describes the formation of multiple images of background sources due to the bending of their light by the mass of intervening objects, can be modeled using score-based priors for the background source \citep[e.g.][]{adam2022cla, Karchev2022}. Such score-based priors have not previously been applied to the problem of jointly inferring the background source and lens mass distribution. Strong lens inversion is particularly challenging in the blind scheme, as the posteriors of parametric lenses generally contain several local minima and exhibit degeneracies between the lens parameters and the source \citep[e.g.][]{Brewer2006, Schneider2013}. Hence, joint inference has only been possible with analytical source priors that impose Gaussian or \textit{smoothness} assumptions \citep[e.g.][]{suyu2006prior, Vegetti2009}.

Strong-lensing observations can enable many sciences, for example measuring $H_0$ via time-delay cosmography \citep[e.g.][]{holycow}, studying high-redshift objects \citep[e.g.][]{Peng2006}, and detecting dark matter subhalos \citep[e.g.][]{Vegetti2012subhalo, Hezaveh2016subhalo}, among other applications. Furthermore, upcoming wide-field surveys, most notably the Rubin Observatory Legacy Survey of Space and Time (LSST) and the \textit{Euclid} space telescope, are expected to observe about 200\,000 strongly lensed systems \citep{Collett2015}.
Advancing strong lens modeling is therefore crucial to extract the full scientific value of this wealth of data. In this paper, we present preliminary results on a new framework for analyzing strong lenses with score-based priors. Our contributions are:
\begin{itemize}
    \item We explore two likelihood score approximations, CLA \citep{adam2022cla} and $\Pi$GDM \citep{song2023pseudoinverseguided}, for source reconstruction.
    \item We adapt GibbsDDRM \citep{murata2023gibbs} to the continuous-time regime, successfully applying it to blind strong-lensing inversion.
    \item We provide empirical evidence that our approach yields residuals consistent with the noise distribution and that the lens marginal posteriors are unbiased.
\end{itemize}
In \Sec{methods} we introduce the method, assumptions, and approximations; in \Sec{experiments} we present experiments on simulated data; and we discuss limitations, outline next steps, and conclude in \Sec{future_limitations} and \Sec{conclusion}.

\begin{figure*}[!htb]
    \centering
    \includegraphics[width=5.5in]{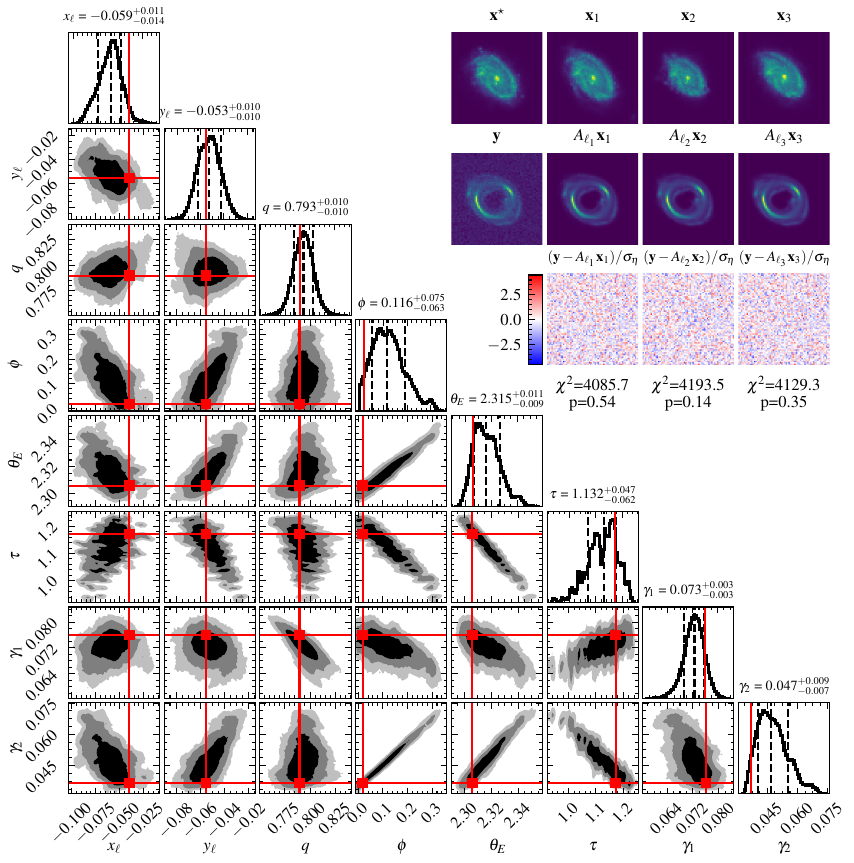}
    \caption{Simulated strong-lensing system analyzed with our joint sampler. \textbf{Top-right panel}: the observed image $\mathbf{y}$, the true source $\mathbf{x}^{\star}$, and three joint-posterior draws $(\mathbf{x}_i,\boldsymbol{\ell}_i)$. For each draw we show the reconstructed image $A_{\boldsymbol{\ell}_i}\mathbf{x}_i$ and the corresponding residual $(\mathbf{y}-A_{\boldsymbol{\ell}_i}\mathbf{x}_i)/\sigma_{\boldsymbol{\eta}}$, demonstrating noise-level consistency. \textbf{Bottom-left panel}: marginal lens posterior $p(\boldsymbol{\ell}\mid\mathbf{y})$ obtained from $406$ joint samples, each augmented with $500$ conditional lens draws as described in \App{marginal_posterior}.}
    \label{fig:main_result}
\end{figure*}

\section{Methods}
\label{sec:methods}

\subsection{Strong gravitational lensing simulations} Strong gravitational lensing can be expressed as a linear operation $\mathbf{y} = A_{\boldsymbol{\ell}}\mathbf{x} + \boldsymbol{\eta}$ ~\citep{Warren2003lenslinear}, where $\mathbf{y}\in \mathbb{R}^m$ is the observed (lensed) image, $\boldsymbol{\ell} \in \mathbb{R}^{n_\ell}$ is the vector of parameters of a parametric lens–mass model, $\mathbf{x}\in \mathbb{R}^n$ is a pixelated representation of the background source, $\boldsymbol{\eta}\sim\mathcal{N}(0,\sigma_{\boldsymbol{\eta}}^2 I)$ is additive Gaussian noise, and $A_{\boldsymbol{\ell}}$ is the Jacobian of the forward model (with dependency on $\boldsymbol{\ell}$ made explicit). Our forward model also includes a point‐spread function (PSF). We use \texttt{Caustics} \citep{stone2024caustic} for the simulations. The lens follows an Elliptical Power‐Law (EPL) profile \citep{Barkana1998epl} with external shear and $m=3$ multipole. The positions of the lens and the source are also free parameters, giving a total of 12 macro parameters besides the pixelated source. See \App{priors} for prior ranges.

\subsection{Score-based models for solving inverse problems}
A generative model for a distribution $p(\mathbf{x})$ can be constructed when we have access to the score $\nabla_{\mathbf{x}_t}\log p_t(\mathbf{x}_t)$ by solving the reverse-time stochastic differential equation (SDE)~\citep{song2021sbm}:
\begin{equation}
    d\mathbf{x}=\bigl[f(\mathbf{x},t)-g(t)^2\,\nabla_{\mathbf{x}}\log p_t(\mathbf{x})\bigr]\,dt
    +g(t)\,d\bar{\mathbf{w}},
\end{equation}
where $p_t(\mathbf{x}_t)$ denotes the target distribution convolved with a perturbation kernel, typically a Gaussian
$\mathcal{N}\!\bigl(\mu(t)\mathbf{x},\sigma(t)^2\mathbb{I}\bigr)$. For the variance-exploding (VE) SDE~\citep{Song2019ve} used in this work,
$\mu(t)=1$ and
$\sigma(t)=\sigma_{\text{min}}\bigl(\sigma_{\text{max}}/\sigma_{\text{min}}\bigr)^{t}$. Given a dataset $\mathcal{D}=\{\mathbf{x}_i\}_{i=1}^N$ with $\mathbf{x}_i\sim p(\mathbf{x})$, we train a neural network $\mathbf{s}_\theta(\mathbf{x}_t,t)$ to approximate the score by minimizing the denoising score-matching loss~\citep{Hyvarinen2005,Vincent2011}:
\begin{equation*}
    \mathcal{L}_\theta=
    \mathbb{E}_{\mathbf{x} \sim \mathcal{D}, t \sim \mathcal{U}(0, 1), \mathbf{x}_t \sim p(\mathbf{x}_t \mid  \mathbf{x})}\bigl[\lambda(t)\,\lVert\mathbf{s}_\theta(\mathbf{x}_t,t)-\nabla_{\mathbf{x}_t}\log p(\mathbf{x}_t\mid\mathbf{x})\rVert^{2}\bigr].
\end{equation*}

\noindent This learns a prior score model from the data examples.  We use \texttt{score-models}\footnote{\href{https://github.com/AlexandreAdam/score\_models}{github.com/AlexandreAdam/score\_models}} to train the network.
Moreover, any score-based model can be turned into a zero-shot posterior sampler \citep[e.g.][]{Graikos2022_plug_and_play} by replacing the prior score with the posterior score:
\begin{equation}
    \nabla_{\mathbf{x}_t}\log p(\mathbf{x}_t\mid\mathbf{y})
    =\nabla_{\mathbf{x}_t}\log p(\mathbf{y}\mid\mathbf{x}_t)
     +\nabla_{\mathbf{x}_t}\log p(\mathbf{x}_t).
\end{equation}
We already have the approximation $\mathbf{s}_\theta(\mathbf{x}_t,t)\approx\nabla_{\mathbf{x}_t}\log p(\mathbf{x}_t)$.  
However, the likelihood score,
\begin{equation}
    \nabla_{\mathbf{x}_t}\log p(\mathbf{y}\mid\mathbf{x}_t)
    =\nabla_{\mathbf{x}_t}\log
      \int_{\mathbf{x}_0}p(\mathbf{x}_0\mid\mathbf{x}_t)\,
      p(\mathbf{y}\mid\mathbf{x}_0)\,d\mathbf{x}_0,
\end{equation}
is intractable.  
Here, we compare three likelihood‐score approximations from the literature across different stages of our inference pipeline: Pseudoinverse-Guided Diffusion Models ($\Pi$GDM)~\citep{song2023pseudoinverseguided}, Convolved Likelihood Approximation (CLA)~\citep{adam2022cla}, and Diffusion Posterior Sampling (DPS)~\citep{chung2023twee}. Using Tweedie’s formula, \citet{chung2023twee} express the posterior mean as
\begin{equation}
\hat{\mathbf{x}}_t\coloneqq\mathbb{E}[\mathbf{x}_0\mid\mathbf{x}_t]
=\mathbf{x}_t+\sigma_t^{2}\,\nabla_{\mathbf{x}_t}\log p(\mathbf{x}_t).
\end{equation}
With $\hat{\mathbf{x}}_t$ in hand, $\Pi$GDM approximates the likelihood as
\begin{equation}
    p_t(\mathbf{y}\mid\mathbf{x}_t)
    \approx
    \mathcal{N}\!\bigl(
        \mathbf{y}\,\bigl|\,
        A_{\boldsymbol{\ell}}\hat{\mathbf{x}}_t,\;
        \Sigma_{\boldsymbol{\eta}}
        +r_t^{2}A_{\boldsymbol{\ell}}A_{\boldsymbol{\ell}}^{\mathsf T}
    \bigr).
\end{equation}
CLA is similar, but it uses $\mathbf{x}_t$ directly in the mean (i.e.\ $A_{\boldsymbol{\ell}}\mathbf{x}_t$) instead of $\hat{\mathbf{x}}_t$. In the original formulation, CLA sets $r_t^{2}=\sigma_t^{2}$, yet \citet{song2023pseudoinverseguided} note that $r_t^{2}$ can, in general, depend on the data and inverse problem. Our choice of $r_t^{2}$ for both approximations is detailed in \App{inference_hyper}.

\subsection{Joint inference of source and lens}
Our goal is to sample from the joint posterior
\begin{equation}
    \mathbf{x},\boldsymbol{\ell}\sim p(\mathbf{x},\boldsymbol{\ell}\mid\mathbf{y})
    \;\propto\;
    \mathcal{N}\!\bigl(\mathbf{y}\mid A_{\boldsymbol{\ell}}\mathbf{x},\Sigma_{\boldsymbol{\eta}}\bigr)\,
    p(\mathbf{x})\,p(\boldsymbol{\ell}),
\end{equation}
where $p(\mathbf{x})$ is implicitly represented by a score-based model, $p(\boldsymbol{\ell})$ is an analytic prior (uniform here), and the likelihood is determined by the noise distribution and the forward model.

We follow GibbsDDRM \citep{murata2023gibbs}, a partially collapsed Gibbs sampler \citep{Dyk2008} that performs Unadjusted Langevin Algorithm (ULA) sampling \citep[e.g.][]{roberts1996ula} of the operator parameters through the reverse diffusion process.

In the original GibbsDDRM formulation, the method builds on DDRM \citep{kawar2022ddrm}, which approximates posterior sampling for non-blind inverse problems. DDRM uses pre-trained DDPM models \citep{Ho2020ddpm} (variance-preserving SDEs in the continuous-time view) re-parameterized in VE notation and conditioned on the observation via the singular-value decomposition of $A$. We instead keep the continuous-time SDE notation with a VE model, integrate with a stochastic Heun solver \citep{Kloeden1992sde}, and use an approximation to the likelihood score at time $t$. Aside from these differences, most procedural details remain the same.

The sampler starts from an initial lens parameter vector $\boldsymbol{\ell}_0$. We find the algorithm to be sensitive to this initialization: the macro-parameters posterior often has multiple local minima, and poor starting points cause the Langevin updates to mix slowly, an issue noted previously in lensing studies \citep[e.g.][]{Brewer2006}. If a given lens has been analyzed before, published values can serve as the initial guess. Alternatively, a learned estimator, such as a CNN \citep[e.g.][]{hezaveh2017cnn,levasseur2017bnn}, can be used.

Here, we obtain $\boldsymbol{\ell}_0$ by minimizing the negative log-likelihood under a Sérsic source model, using the Adam optimizer \citep{kingma2017adam} with a multi-start strategy. Given $\boldsymbol{\ell}_0$, we solve the reverse-time SDE. At diffusion time~$t$, the source $\mathbf{x}_t$ is updated while keeping $\boldsymbol{\ell}$ fixed, employing $\Pi$GDM for the likelihood score with respect to $\mathbf{x}_t$ (see \App{inference_hyper} for a complete justification for using $\Pi$GDM over CLA). Next, we update the lens parameters via ULA using
\begin{equation}
    \nabla_{\boldsymbol{\ell}}\log p(\boldsymbol{\ell}\mid\mathbf{y},\mathbf{x}_t)
    \;\approx\;
    \nabla_{\boldsymbol{\ell}}
    \log\bigl[\mathcal{N}(\mathbf{y}\mid
    A_{\boldsymbol{\ell}}\hat{\mathbf{x}}_t,\Sigma_{\boldsymbol{\eta}})\,p(\boldsymbol{\ell})\bigr].
\end{equation}
In summary, we run a partially collapsed Gibbs sampler through the reverse diffusion process, alternating between updating the lens parameters with ULA and the source with $\Pi$GDM. Finally, we perform a full Gibbs sweep starting from the joint sample
$(\mathbf{x},\boldsymbol{\ell})\sim p(\mathbf{x},\boldsymbol{\ell}\mid\mathbf{y})$ updating the source with CLA, which has been more tested for non-blind lensing inversion,
and the lens with NUTS \citep{Hoffman2014nuts}.

\section{Experiments and results}
\label{sec:experiments}

\subsection{Dataset}
We use $60\,774$ images from the \textsc{SKIRT}--TNG dataset \citep{Bottrell2024}, produced by applying dust–radiative‐transfer post-processing \citep{Camps2020} to galaxies in the TNG cosmological magneto-hydrodynamical simulations\footnote{\url{www.tng-project.org}} \citep{Nelson2019}.
The $i$-band frames are downsampled to $64 \times 64$ pixels and converted to flux units of $\mu\mathrm{Jy}\,\mathrm{sr}^{-1}$ to train the SBM. A further $4\,293$ images from the same dataset are reserved as a validation set, used only for inference.

\subsection{Joint–inference experiment}
\Fig{main_result} illustrates a representative run of our joint sampler.  
The observation $\mathbf{y}$ is generated from a ground-truth source $\mathbf{x}^{\star}$ and lens parameters $\boldsymbol{\ell}^{\star}$.  
In the top right panel, we display three joint-posterior draws $(\mathbf{x}_{i},\boldsymbol{\ell}_{i}) \sim p(\mathbf{x},\boldsymbol{\ell} \mid \mathbf{y})$.  
Each sampled source recovers the overall morphology of $\mathbf{x}^{\star}$ while exhibiting natural variability in size and pixel-scale structure.  
The corresponding reconstructions $A_{\boldsymbol{\ell}_{i}}\mathbf{x}_{i}$ closely match the observation, and the normalized residuals are consistent with the noise model. The bottom-left panel shows the marginal lens posterior $p(\boldsymbol{\ell} \mid \mathbf{y})$, estimated from $406$ joint samples as detailed in \App{marginal_posterior}.  
Known degeneracies among strong-lensing parameters are properly explored, and the true lens parameters $\boldsymbol{\ell}^{\star}$ lie well within the high-probability region.  
A full corner plot of all $12$ lens parameters, additional source–reconstruction pairs, and three further lensing systems are provided in \App{extra_experiments}.

\section{Future work and limitations}
\label{sec:future_limitations}

Although the results are promising, we still need to test the robustness and generality of the method. In future iterations, we plan to run a coverage test with TARP \citep{lemos2023tarp}, a sample-based diagnostic that is necessary and sufficient for posterior coverage. Furthermore, the overall pipeline contains several hyperparameters and design choices that remain unexplored. For example, we intend to perform an ablation study on the $r_t^{2}$ parameter in both CLA and \(\Pi\)GDM for this specific problem. We also plan to benchmark non-blind inverse approaches in lensing, following the protocol of \citet{zheng2025inversebench}.

A key limitation of our approach is its strong dependence on the initialization \(\boldsymbol{\ell}_0\). Because ULA explores a single mode and implicitly assumes a log-concave target, multi-modal or strongly non-log-concave posteriors are difficult to sample. Preliminary experiments show that mode switching (or recovery from a poor initialization) is possible when using a \emph{swarm} of walkers, or by performing expectation–maximization through the diffusion process instead of sampling, as in \citet{laroche2024fastem}.

Another limitation of the current framework is its assumption of a purely additive Gaussian noise model with a known likelihood. This may not fully capture the complex noise properties of real observations. Future work will explore relaxing this assumption. One direction is to adopt methods that can handle complex or non-Gaussian noise, for instance, by building a data-driven noise model using a score-based likelihood characterization \citep[e.g.][]{Legin2023slic}. A second, complementary approach is to treat observational uncertainty as an intrinsic part of the diffusion process itself, using deconvolution or mixed-noise learning techniques \citep[e.g.][]{Lu2025noisy, Hagemann2025mixed}.

Finally, it would be valuable to compare analytical priors with learned priors when the score is available. A natural baseline is a Gaussian prior, which has a long history in lensing modeling \citep[e.g.][]{suyu2006prior}. The full source code implementing our sampler will be released publicly upon journal publication.

\section{Conclusion}
\label{sec:conclusion}

We have presented the first score-based framework for \emph{blind} strong-lensing inversion, jointly sampling the pixelated source and the parametric lens mass.  Our approach couples a VE score model with a GibbsDDRM-like sampler: the source is updated with $\Pi$GDM, and the lens parameters with an Unadjusted Langevin step. On simulated data, the sampler yields residuals that match the noise distribution and recovers all lens parameters without systematic bias, showing that it can handle the nonlinearities and parameter degeneracies characteristic of strong lensing. In future work, we plan to improve and extend the framework and apply it to the large samples of lenses expected from LSST and \textit{Euclid}, enabling fully Bayesian analyses at next-generation survey scale.

\begin{ack}
This work is partially supported by Schmidt Futures, a philanthropic initiative founded by Eric and Wendy Schmidt as part of the Virtual Institute for Astrophysics (VIA). The work is in part supported by computational resources provided by Calcul Québec and the Digital Research Alliance of Canada. G.M.B. thanks Sacha Perry-Fagant, Nicolas Payot, and Alexandre Adam for useful discussions and valuable input while this work was being completed. Y.H. and L.P.-L. acknowledge support from the Canada Research Chairs Program, the National Sciences and Engineering Council of Canada (NSERC) through grants RGPIN-2020-05073 and 05102, and the Fonds de recherche du Québec through grant CFQCU-2024-348060. R.L. acknowledges support from the NSERC CGS D scholarship. C.S. acknowledges the support of an NSERC Postdoctoral Fellowship and a CITA National Fellowship. G.M.B. acknowledges support from the Fonds de recherche du Québec – Nature et technologies (FRQNT) under a Doctoral Research Scholarship (\doi{10.69777/368273}).
\end{ack}

\bibliographystyle{unsrtnat}
\bibliography{bib}

\providecommand{\apj}{The Astrophysical Journal} \providecommand{\mnras}{Monthly Notices of the Royal Astronomical Society} \providecommand{\apjl}{The Astrophysical Journal Letter} \providecommand{\nat}{Nature}
\begin{thebibliography}{60}
\providecommand{\natexlab}[1]{#1}
\providecommand{\url}[1]{\texttt{#1}}
\expandafter\ifx\csname urlstyle\endcsname\relax
  \providecommand{\doi}[1]{doi: #1}\else
  \providecommand{\doi}{doi: \begingroup \urlstyle{rm}\Url}\fi

\bibitem[Feng et~al.(2023)Feng, Smith, Rubinstein, Chang, Bouman, and Freeman]{feng2023score}
Berthy~T. Feng, Jamie Smith, Michael Rubinstein, Huiwen Chang, Katherine~L. Bouman, and William~T. Freeman.
\newblock Score-based diffusion models as principled priors for inverse imaging.
\newblock In \emph{International Conference on Computer Vision (ICCV)}, pages 10486--10497, 2023.
\newblock URL \url{https://doi.org/10.1109/ICCV51070.2023.00965}.

\bibitem[Dia et~al.(2025)Dia, Yantovski-Barth, Adam, Bowles, Perreault-Levasseur, Hezaveh, and Scaife]{dia2025iris}
Noé Dia, M.~J. Yantovski-Barth, Alexandre Adam, Micah Bowles, Laurence Perreault-Levasseur, Yashar Hezaveh, and Anna Scaife.
\newblock Iris: A bayesian approach for image reconstruction in radio interferometry with expressive score-based priors, 2025.
\newblock URL \url{https://arxiv.org/abs/2501.02473}.

\bibitem[{Adam} et~al.(2022){Adam}, {Coogan}, {Malkin}, {Legin}, {Perreault-Levasseur}, {Hezaveh}, and {Bengio}]{adam2022cla}
Alexandre {Adam}, Adam {Coogan}, Nikolay {Malkin}, Ronan {Legin}, Laurence {Perreault-Levasseur}, Yashar {Hezaveh}, and Yoshua {Bengio}.
\newblock {Posterior samples of source galaxies in strong gravitational lenses with score-based priors}.
\newblock In \emph{Machine Learning and the Physical Sciences Workshop, NeurIPS 2022}, January 2022.
\newblock URL \url{https://arxiv.org/pdf/2211.03812}.

\bibitem[Karchev et~al.(2022)Karchev, Anau~Montel, Coogan, and Weniger]{Karchev2022}
Konstantin Karchev, Noemi Anau~Montel, Adam Coogan, and Christoph Weniger.
\newblock {Strong-Lensing Source Reconstruction with Denoising Diffusion Restoration Models}.
\newblock In \emph{Machine Learning and the Physical Sciences Workshop, NeurIPS 2022}, 11 2022.
\newblock URL \url{https://arxiv.org/pdf/2211.04365}.

\bibitem[Legin et~al.(2023)Legin, Ho, Lemos, Perreault-Levasseur, Ho, Hezaveh, and Wandelt]{Legin2023}
Ronan Legin, Matthew Ho, Pablo Lemos, Laurence Perreault-Levasseur, Shirley Ho, Yashar Hezaveh, and Benjamin Wandelt.
\newblock Posterior sampling of the initial conditions of the universe from non-linear large scale structures using score-based generative models.
\newblock \emph{Monthly Notices of the Royal Astronomical Society: Letters}, 527\penalty0 (1):\penalty0 L173--L178, 10 2023.
\newblock ISSN 1745-3925.
\newblock \doi{10.1093/mnrasl/slad152}.

\bibitem[{Ono} et~al.(2024){Ono}, {Park}, {Mudur}, {Ni}, {Cuesta-Lazaro}, and {Villaescusa-Navarro}]{Ono2024}
Victoria {Ono}, Core~Francisco {Park}, Nayantara {Mudur}, Yueying {Ni}, Carolina {Cuesta-Lazaro}, and Francisco {Villaescusa-Navarro}.
\newblock {Debiasing with Diffusion: Probabilistic Reconstruction of Dark Matter Fields from Galaxies with CAMELS}.
\newblock \emph{The Astrophysical Journal}, 970\penalty0 (2):\penalty0 174, August 2024.
\newblock \doi{10.3847/1538-4357/ad5957}.

\bibitem[{Fl{\"o}ss} et~al.(2024){Fl{\"o}ss}, {Coulton}, {Duivenvoorden}, {Villaescusa-Navarro}, and {Wandelt}]{Floss2024}
Thomas {Fl{\"o}ss}, William~R. {Coulton}, Adriaan~J. {Duivenvoorden}, Francisco {Villaescusa-Navarro}, and Benjamin~D. {Wandelt}.
\newblock {Denoising diffusion delensing: reconstructing the non-Gaussian CMB lensing potential with diffusion models}.
\newblock \emph{Monthly Notices of the Royal Astronomical Society}, 533\penalty0 (1):\penalty0 423--432, September 2024.
\newblock \doi{10.1093/mnras/stae1818}.

\bibitem[Xue et~al.(2023)Xue, Li, Patel, and Regier]{xue2023}
Zhiwei Xue, Yuhang Li, Yash Patel, and Jeffrey Regier.
\newblock Diffusion models for probabilistic deconvolution of galaxy images.
\newblock In \emph{Machine Learning for Astrophysics Workshop, ICML 2023}, 2023.
\newblock URL \url{https://arxiv.org/abs/2307.11122}.

\bibitem[{Adam} et~al.(2025){Adam}, {Stone}, {Bottrell}, {Legin}, {Hezaveh}, and {Perreaul-Levasseur}]{Adam2025}
Alexandre {Adam}, Connor {Stone}, Connor {Bottrell}, Ronan {Legin}, Yashar {Hezaveh}, and Laurence {Perreaul-Levasseur}.
\newblock {Echoes in the Noise: Posterior Samples of Faint Galaxy Surface Brightness Profiles with Score-based Likelihoods and Priors}.
\newblock \emph{The Astronomical Journal}, 169\penalty0 (5):\penalty0 254, May 2025.
\newblock \doi{10.3847/1538-3881/adb039}.

\bibitem[Zheng et~al.(2025)Zheng, Chu, Zhang, Wu, Wang, Feng, Zou, Sun, Kovachki, Ross, Bouman, and Yue]{zheng2025inversebench}
Hongkai Zheng, Wenda Chu, Bingliang Zhang, Zihui Wu, Austin Wang, Berthy Feng, Caifeng Zou, Yu~Sun, Nikola~Borislavov Kovachki, Zachary~E Ross, Katherine Bouman, and Yisong Yue.
\newblock Inversebench: Benchmarking plug-and-play diffusion priors for inverse problems in physical sciences.
\newblock In \emph{The Thirteenth International Conference on Learning Representations}, 2025.
\newblock URL \url{https://openreview.net/forum?id=U3PBITXNG6}.

\bibitem[Song et~al.(2023)Song, Vahdat, Mardani, and Kautz]{song2023pseudoinverseguided}
Jiaming Song, Arash Vahdat, Morteza Mardani, and Jan Kautz.
\newblock Pseudoinverse-guided diffusion models for inverse problems.
\newblock In \emph{The Eleventh International Conference on Learning Representations}, 2023.
\newblock URL \url{https://openreview.net/forum?id=9_gsMA8MRKQ}.

\bibitem[Chung et~al.(2023{\natexlab{a}})Chung, Kim, Mccann, Klasky, and Ye]{chung2023twee}
Hyungjin Chung, Jeongsol Kim, Michael~Thompson Mccann, Marc~Louis Klasky, and Jong~Chul Ye.
\newblock Diffusion posterior sampling for general noisy inverse problems.
\newblock In \emph{The Eleventh International Conference on Learning Representations}, 2023{\natexlab{a}}.
\newblock URL \url{https://openreview.net/forum?id=OnD9zGAGT0k}.

\bibitem[Wu et~al.(2024)Wu, Sun, Chen, Zhang, Yue, and Bouman]{Wu2024}
Zihui Wu, Yu~Sun, Yifan Chen, Bingliang Zhang, Yisong Yue, and Katherine~L. Bouman.
\newblock Principled probabilistic imaging using diffusion models as plug-and-play priors.
\newblock In A.~Globerson, L.~Mackey, D.~Belgrave, A.~Fan, U.~Paquet, J.~Tomczak, and C.~Zhang, editors, \emph{Advances in Neural Information Processing Systems}, volume~37, pages 118389--118427. Curran Associates, Inc., 2024.
\newblock URL \url{https://proceedings.neurips.cc/paper_files/paper/2024/file/d65c4ce22241138c1784ff753d4c746c-Paper-Conference.pdf}.

\bibitem[Feng and Bouman(2024)]{feng2024variational}
Berthy Feng and Katherine Bouman.
\newblock Variational bayesian imaging with an efficient surrogate score-based prior.
\newblock \emph{Transactions on Machine Learning Research}, 2024.
\newblock ISSN 2835-8856.
\newblock URL \url{https://openreview.net/forum?id=db2pFKVcm1}.

\bibitem[Dou and Song(2023)]{Dou2023}
Zehao Dou and Yang Song.
\newblock Diffusion {Posterior} {Sampling} for {Linear} {Inverse} {Problem} {Solving}: {A} {Filtering} {Perspective}.
\newblock In \emph{The Twelfth International Conference on Learning Representations}, October 2023.
\newblock URL \url{https://openreview.net/forum?id=tplXNcHZs1}.

\bibitem[Levin et~al.(2009)Levin, Weiss, Durand, and Freeman]{Levin2009blind}
Anat Levin, Yair Weiss, Fredo Durand, and William~T. Freeman.
\newblock Understanding and evaluating blind deconvolution algorithms.
\newblock In \emph{2009 IEEE Conference on Computer Vision and Pattern Recognition}, pages 1964--1971, 2009.
\newblock \doi{10.1109/CVPR.2009.5206815}.

\bibitem[Gao et~al.(2021)Gao, Castellanos, Yue, Ross, and Bouman]{Gao2021blind_gem}
Angela Gao, Jorge Castellanos, Yisong Yue, Zachary Ross, and Katherine Bouman.
\newblock Deepgem: Generalized expectation-maximization for blind inversion.
\newblock In M.~Ranzato, A.~Beygelzimer, Y.~Dauphin, P.S. Liang, and J.~Wortman Vaughan, editors, \emph{Advances in Neural Information Processing Systems}, volume~34, pages 11592--11603. Curran Associates, Inc., 2021.
\newblock URL \url{https://proceedings.neurips.cc/paper_files/paper/2021/file/606c90a06173d69682feb83037a68fec-Paper.pdf}.

\bibitem[Murata et~al.(2023)Murata, Saito, Lai, Takida, Uesaka, Mitsufuji, and Ermon]{murata2023gibbs}
Naoki Murata, Koichi Saito, Chieh-Hsin Lai, Yuhta Takida, Toshimitsu Uesaka, Yuki Mitsufuji, and Stefano Ermon.
\newblock {G}ibbs{DDRM}: A partially collapsed {G}ibbs sampler for solving blind inverse problems with denoising diffusion restoration.
\newblock In Andreas Krause, Emma Brunskill, Kyunghyun Cho, Barbara Engelhardt, Sivan Sabato, and Jonathan Scarlett, editors, \emph{Proceedings of the 40th International Conference on Machine Learning}, volume 202 of \emph{Proceedings of Machine Learning Research}, pages 25501--25522. PMLR, 23--29 Jul 2023.
\newblock URL \url{https://proceedings.mlr.press/v202/murata23a.html}.

\bibitem[Chung et~al.(2023{\natexlab{b}})Chung, Kim, Kim, and Ye]{chung2023blinddps}
Hyungjin Chung, Jeongsol Kim, Sehui Kim, and Jong~Chul Ye.
\newblock Parallel {Diffusion} {Models} of {Operator} and {Image} for {Blind} {Inverse} {Problems}.
\newblock In \emph{2023 {IEEE}/{CVF} {Conference} on {Computer} {Vision} and {Pattern} {Recognition} ({CVPR})}, pages 6059--6069, Vancouver, BC, Canada, June 2023{\natexlab{b}}. IEEE.
\newblock ISBN 979-8-3503-0129-8.
\newblock \doi{10.1109/CVPR52729.2023.00587}.

\bibitem[Chihaoui et~al.(2024)Chihaoui, Lemkhenter, and Favaro]{chihaoui2024bird}
Hamadi Chihaoui, Abdelhak Lemkhenter, and Paolo Favaro.
\newblock Blind image restoration via fast diffusion inversion.
\newblock In \emph{The Thirty-eighth Annual Conference on Neural Information Processing Systems}, 2024.
\newblock URL \url{https://openreview.net/forum?id=HfSJlBRkKJ}.

\bibitem[Laroche et~al.(2024)Laroche, Almansa, and Coupete]{laroche2024fastem}
Charles Laroche, Andrés Almansa, and Eva Coupete.
\newblock Fast {Diffusion} {EM}: a diffusion model for blind inverse problems with application to deconvolution.
\newblock In \emph{2024 {IEEE}/{CVF} {Winter} {Conference} on {Applications} of {Computer} {Vision} ({WACV})}, pages 5259--5269, Waikoloa, HI, USA, January 2024. IEEE.
\newblock ISBN 979-8-3503-1892-0.
\newblock \doi{10.1109/WACV57701.2024.00519}.

\bibitem[{Brewer} and {Lewis}(2006)]{Brewer2006}
Brendon~J. {Brewer} and Geraint~F. {Lewis}.
\newblock {Strong Gravitational Lens Inversion: A Bayesian Approach}.
\newblock \emph{The Astrophysical Journal}, 637\penalty0 (2):\penalty0 608--619, February 2006.
\newblock \doi{10.1086/498409}.

\bibitem[{Schneider} and {Sluse}(2013)]{Schneider2013}
Peter {Schneider} and Dominique {Sluse}.
\newblock {Mass-sheet degeneracy, power-law models and external convergence: Impact on the determination of the Hubble constant from gravitational lensing}.
\newblock \emph{Astronomy and Astrophysics}, 559:\penalty0 A37, November 2013.
\newblock \doi{10.1051/0004-6361/201321882}.

\bibitem[{Suyu} et~al.(2006){Suyu}, {Marshall}, {Hobson}, and {Blandford}]{suyu2006prior}
S.~H. {Suyu}, P.~J. {Marshall}, M.~P. {Hobson}, and R.~D. {Blandford}.
\newblock {A Bayesian analysis of regularized source inversions in gravitational lensing}.
\newblock \emph{\mnras}, 371\penalty0 (2):\penalty0 983--998, September 2006.
\newblock \doi{10.1111/j.1365-2966.2006.10733.x}.

\bibitem[{Vegetti} and {Koopmans}(2009)]{Vegetti2009}
S.~{Vegetti} and L.~V.~E. {Koopmans}.
\newblock {Bayesian strong gravitational-lens modelling on adaptive grids: objective detection of mass substructure in Galaxies}.
\newblock \emph{\mnras}, 392\penalty0 (3):\penalty0 945--963, January 2009.
\newblock \doi{10.1111/j.1365-2966.2008.14005.x}.

\bibitem[{Wong} et~al.(2020){Wong}, {Suyu}, {Chen}, {Rusu}, {Million}, {Sluse}, {Bonvin}, {Fassnacht}, {Taubenberger}, {Auger}, {Birrer}, {Chan}, {Courbin}, {Hilbert}, {Tihhonova}, {Treu}, {Agnello}, {Ding}, {Jee}, {Komatsu}, {Shajib}, {Sonnenfeld}, {Blandford}, {Koopmans}, {Marshall}, and {Meylan}]{holycow}
Kenneth~C. {Wong}, Sherry~H. {Suyu}, Geoff C.~F. {Chen}, Cristian~E. {Rusu}, Martin {Million}, Dominique {Sluse}, Vivien {Bonvin}, Christopher~D. {Fassnacht}, Stefan {Taubenberger}, Matthew~W. {Auger}, Simon {Birrer}, James H.~H. {Chan}, Frederic {Courbin}, Stefan {Hilbert}, Olga {Tihhonova}, Tommaso {Treu}, Adriano {Agnello}, Xuheng {Ding}, In~{Jee}, Eiichiro {Komatsu}, Anowar~J. {Shajib}, Alessandro {Sonnenfeld}, Roger~D. {Blandford}, L{\'e}on V.~E. {Koopmans}, Philip~J. {Marshall}, and Georges {Meylan}.
\newblock {H0LiCOW - XIII. A 2.4 per cent measurement of H$_{0}$ from lensed quasars: 5.3{\ensuremath{\sigma}} tension between early- and late-Universe probes}.
\newblock \emph{MNRAS}, 498\penalty0 (1):\penalty0 1420--1439, October 2020.
\newblock \doi{10.1093/mnras/stz3094}.

\bibitem[{Peng} et~al.(2006){Peng}, {Impey}, {Rix}, {Kochanek}, {Keeton}, {Falco}, {Leh{\'a}r}, and {McLeod}]{Peng2006}
Chien~Y. {Peng}, Chris~D. {Impey}, Hans-Walter {Rix}, Christopher~S. {Kochanek}, Charles~R. {Keeton}, Emilio~E. {Falco}, Joseph {Leh{\'a}r}, and Brian~A. {McLeod}.
\newblock {Probing the Coevolution of Supermassive Black Holes and Galaxies Using Gravitationally Lensed Quasar Hosts}.
\newblock \emph{ApJ}, 649\penalty0 (2):\penalty0 616--634, October 2006.
\newblock \doi{10.1086/506266}.

\bibitem[{Vegetti} et~al.(2012){Vegetti}, {Lagattuta}, {McKean}, {Auger}, {Fassnacht}, and {Koopmans}]{Vegetti2012subhalo}
S.~{Vegetti}, D.~J. {Lagattuta}, J.~P. {McKean}, M.~W. {Auger}, C.~D. {Fassnacht}, and L.~V.~E. {Koopmans}.
\newblock {Gravitational detection of a low-mass dark satellite galaxy at cosmological distance}.
\newblock \emph{Nature}, 481\penalty0 (7381):\penalty0 341--343, January 2012.
\newblock \doi{10.1038/nature10669}.

\bibitem[{Hezaveh} et~al.(2016){Hezaveh}, {Dalal}, {Marrone}, {Mao}, {Morningstar}, {Wen}, {Blandford}, {Carlstrom}, {Fassnacht}, {Holder}, {Kemball}, {Marshall}, {Murray}, {Perreault Levasseur}, {Vieira}, and {Wechsler}]{Hezaveh2016subhalo}
Yashar~D. {Hezaveh}, Neal {Dalal}, Daniel~P. {Marrone}, Yao-Yuan {Mao}, Warren {Morningstar}, Di~{Wen}, Roger~D. {Blandford}, John~E. {Carlstrom}, Christopher~D. {Fassnacht}, Gilbert~P. {Holder}, Athol {Kemball}, Philip~J. {Marshall}, Norman {Murray}, Laurence {Perreault Levasseur}, Joaquin~D. {Vieira}, and Risa~H. {Wechsler}.
\newblock {Detection of Lensing Substructure Using ALMA Observations of the Dusty Galaxy SDP.81}.
\newblock \emph{The Astrophysical Journal}, 823\penalty0 (1):\penalty0 37, May 2016.
\newblock \doi{10.3847/0004-637X/823/1/37}.

\bibitem[{Collett}(2015)]{Collett2015}
Thomas~E. {Collett}.
\newblock {The Population of Galaxy-Galaxy Strong Lenses in Forthcoming Optical Imaging Surveys}.
\newblock \emph{The Astrophysical Journal}, 811\penalty0 (1):\penalty0 20, September 2015.
\newblock \doi{10.1088/0004-637X/811/1/20}.

\bibitem[{Warren} and {Dye}(2003)]{Warren2003lenslinear}
S.~J. {Warren} and S.~{Dye}.
\newblock {Semilinear Gravitational Lens Inversion}.
\newblock \emph{The Astrophysical Journal}, 590\penalty0 (2):\penalty0 673--682, June 2003.
\newblock \doi{10.1086/375132}.

\bibitem[{Stone} et~al.(2024){Stone}, {Adam}, {Coogan}, {Yantovski-Barth}, {Filipp}, {Setiawan}, {Core}, {Legin}, {Wilson}, {Barco}, {Hezaveh}, and {Perreault-Levasseur}]{stone2024caustic}
Connor {Stone}, Alexandre {Adam}, Adam {Coogan}, M.~{Yantovski-Barth}, Andreas {Filipp}, Landung {Setiawan}, Cordero {Core}, Ronan {Legin}, Charles {Wilson}, Gabriel {Barco}, Yashar {Hezaveh}, and Laurence {Perreault-Levasseur}.
\newblock {Caustics: A Python Package for Accelerated Strong Gravitational Lensing Simulations}.
\newblock \emph{The Journal of Open Source Software}, 9\penalty0 (103):\penalty0 7081, November 2024.
\newblock \doi{10.21105/joss.07081}.

\bibitem[{Barkana}(1998)]{Barkana1998epl}
Rennan {Barkana}.
\newblock {Fast Calculation of a Family of Elliptical Mass Gravitational Lens Models}.
\newblock \emph{The Astrophysical Journal}, 502\penalty0 (2):\penalty0 531--537, August 1998.
\newblock \doi{10.1086/305950}.

\bibitem[Song et~al.(2021)Song, Sohl-Dickstein, Kingma, Kumar, Ermon, and Poole]{song2021sbm}
Yang Song, Jascha Sohl-Dickstein, Diederik~P Kingma, Abhishek Kumar, Stefano Ermon, and Ben Poole.
\newblock Score-based generative modeling through stochastic differential equations.
\newblock In \emph{The Ninth International Conference on Learning Representations}, 2021.
\newblock URL \url{https://openreview.net/forum?id=PxTIG12RRHS}.

\bibitem[Song and Ermon(2019)]{Song2019ve}
Yang Song and Stefano Ermon.
\newblock Generative modeling by estimating gradients of the data distribution.
\newblock In H.~Wallach, H.~Larochelle, A.~Beygelzimer, F.~d\textquotesingle Alch\'{e}-Buc, E.~Fox, and R.~Garnett, editors, \emph{Advances in Neural Information Processing Systems}, volume~32. Curran Associates, Inc., 2019.
\newblock URL \url{https://proceedings.neurips.cc/paper_files/paper/2019/file/3001ef257407d5a371a96dcd947c7d93-Paper.pdf}.

\bibitem[Hyv{{\"a}}rinen(2005)]{Hyvarinen2005}
Aapo Hyv{{\"a}}rinen.
\newblock Estimation of non-normalized statistical models by score matching.
\newblock \emph{Journal of Machine Learning Research}, 6\penalty0 (24):\penalty0 695--709, 2005.
\newblock URL \url{http://jmlr.org/papers/v6/hyvarinen05a.html}.

\bibitem[Vincent(2011)]{Vincent2011}
Pascal Vincent.
\newblock A connection between score matching and denoising autoencoders.
\newblock \emph{Neural Comput.}, 23\penalty0 (7):\penalty0 1661--1674, 2011.
\newblock \doi{10.1162/NECO\_a\_00142}.

\bibitem[Graikos et~al.(2022)Graikos, Malkin, Jojic, and Samaras]{Graikos2022_plug_and_play}
Alexandros Graikos, Nikolay Malkin, Nebojsa Jojic, and Dimitris Samaras.
\newblock Diffusion models as plug-and-play priors.
\newblock In S.~Koyejo, S.~Mohamed, A.~Agarwal, D.~Belgrave, K.~Cho, and A.~Oh, editors, \emph{Advances in Neural Information Processing Systems}, volume~35, pages 14715--14728. Curran Associates, Inc., 2022.
\newblock URL \url{https://proceedings.neurips.cc/paper_files/paper/2022/file/5e6cec2a9520708381fe520246018e8b-Paper-Conference.pdf}.

\bibitem[Van~Dyk and Park(2008)]{Dyk2008}
David~A Van~Dyk and Taeyoung Park.
\newblock Partially collapsed gibbs samplers.
\newblock \emph{Journal of the American Statistical Association}, 103\penalty0 (482):\penalty0 790--796, 2008.
\newblock \doi{10.1198/016214508000000409}.

\bibitem[Roberts and Tweedie(1996)]{roberts1996ula}
Gareth~O. Roberts and Richard~L. Tweedie.
\newblock Exponential convergence of {Langevin} distributions and their discrete approximations.
\newblock \emph{Bernoulli}, 2\penalty0 (4):\penalty0 341--363, December 1996.
\newblock ISSN 1350-7265.
\newblock URL \url{https://projecteuclid.org/journals/bernoulli/volume-2/issue-4/Exponential-convergence-of-Langevin-distributions-and-their-discrete-approximations/bj/1178291835.full}.
\newblock Publisher: Bernoulli Society for Mathematical Statistics and Probability.

\bibitem[Kawar et~al.(2022)Kawar, Elad, Ermon, and Song]{kawar2022ddrm}
Bahjat Kawar, Michael Elad, Stefano Ermon, and Jiaming Song.
\newblock Denoising diffusion restoration models.
\newblock In S.~Koyejo, S.~Mohamed, A.~Agarwal, D.~Belgrave, K.~Cho, and A.~Oh, editors, \emph{Advances in Neural Information Processing Systems}, volume~35, pages 23593--23606. Curran Associates, Inc., 2022.
\newblock URL \url{https://proceedings.neurips.cc/paper_files/paper/2022/file/95504595b6169131b6ed6cd72eb05616-Paper-Conference.pdf}.

\bibitem[Ho et~al.(2020)Ho, Jain, and Abbeel]{Ho2020ddpm}
Jonathan Ho, Ajay Jain, and Pieter Abbeel.
\newblock Denoising diffusion probabilistic models.
\newblock In H.~Larochelle, M.~Ranzato, R.~Hadsell, M.F. Balcan, and H.~Lin, editors, \emph{Advances in Neural Information Processing Systems}, volume~33, pages 6840--6851. Curran Associates, Inc., 2020.
\newblock URL \url{https://proceedings.neurips.cc/paper_files/paper/2020/file/4c5bcfec8584af0d967f1ab10179ca4b-Paper.pdf}.

\bibitem[Kloeden and Platen(1992)]{Kloeden1992sde}
Peter~E. Kloeden and Eckhard Platen.
\newblock \emph{Numerical {Solution} of {Stochastic} {Differential} {Equations}}.
\newblock Springer, Berlin, Heidelberg, 1992.
\newblock ISBN 978-3-642-08107-1 978-3-662-12616-5.
\newblock \doi{10.1007/978-3-662-12616-5}.

\bibitem[{Hezaveh} et~al.(2017){Hezaveh}, {Perreault Levasseur}, and {Marshall}]{hezaveh2017cnn}
Yashar~D. {Hezaveh}, Laurence {Perreault Levasseur}, and Philip~J. {Marshall}.
\newblock {Fast automated analysis of strong gravitational lenses with convolutional neural networks}.
\newblock \emph{\nat}, 548\penalty0 (7669):\penalty0 555--557, August 2017.
\newblock \doi{10.1038/nature23463}.

\bibitem[{Perreault Levasseur} et~al.(2017){Perreault Levasseur}, {Hezaveh}, and {Wechsler}]{levasseur2017bnn}
Laurence {Perreault Levasseur}, Yashar~D. {Hezaveh}, and Risa~H. {Wechsler}.
\newblock {Uncertainties in Parameters Estimated with Neural Networks: Application to Strong Gravitational Lensing}.
\newblock \emph{\apjl}, 850\penalty0 (1):\penalty0 L7, November 2017.
\newblock \doi{10.3847/2041-8213/aa9704}.

\bibitem[Kingma and Ba(2015)]{kingma2017adam}
Diederik~P. Kingma and Jimmy Ba.
\newblock Adam: {A} method for stochastic optimization.
\newblock In Yoshua Bengio and Yann LeCun, editors, \emph{3rd International Conference on Learning Representations, {ICLR} 2015, San Diego, CA, USA, May 7-9, 2015, Conference Track Proceedings}, 2015.
\newblock URL \url{http://arxiv.org/abs/1412.6980}.

\bibitem[Hoffman and Gelman(2014)]{Hoffman2014nuts}
Matthew~D. Hoffman and Andrew Gelman.
\newblock The no-u-turn sampler: Adaptively setting path lengths in hamiltonian monte carlo.
\newblock \emph{Journal of Machine Learning Research}, 15\penalty0 (47):\penalty0 1593--1623, 2014.
\newblock URL \url{http://jmlr.org/papers/v15/hoffman14a.html}.

\bibitem[{Bottrell} et~al.(2024){Bottrell}, {Yesuf}, {Popping}, {Omori}, {Tang}, {Ding}, {Pillepich}, {Nelson}, {Eisert}, {Gao}, {Goulding}, {Kalita}, {Luo}, {Greene}, {Shi}, and {Silverman}]{Bottrell2024}
Connor {Bottrell}, Hassen~M. {Yesuf}, Gerg{\"o} {Popping}, Kiyoaki~Christopher {Omori}, Shenli {Tang}, Xuheng {Ding}, Annalisa {Pillepich}, Dylan {Nelson}, Lukas {Eisert}, Hua {Gao}, Andy~D. {Goulding}, Boris~S. {Kalita}, Wentao {Luo}, Jenny~E. {Greene}, Jingjing {Shi}, and John~D. {Silverman}.
\newblock {IllustrisTNG in the HSC-SSP: image data release and the major role of mini mergers as drivers of asymmetry and star formation}.
\newblock \emph{MNRAS}, 527\penalty0 (3):\penalty0 6506--6539, January 2024.
\newblock \doi{10.1093/mnras/stad2971}.

\bibitem[Camps and Baes(2020)]{Camps2020}
P.~Camps and M.~Baes.
\newblock Skirt 9: Redesigning an advanced dust radiative transfer code to allow kinematics, line transfer and polarization by aligned dust grains.
\newblock \emph{Astronomy and Computing}, 31:\penalty0 100381, 2020.
\newblock ISSN 2213-1337.
\newblock \doi{10.1016/j.ascom.2020.100381}.

\bibitem[{Nelson} et~al.(2019){Nelson}, {Springel}, {Pillepich}, {Rodriguez-Gomez}, {Torrey}, {Genel}, {Vogelsberger}, {Pakmor}, {Marinacci}, {Weinberger}, {Kelley}, {Lovell}, {Diemer}, and {Hernquist}]{Nelson2019}
Dylan {Nelson}, Volker {Springel}, Annalisa {Pillepich}, Vicente {Rodriguez-Gomez}, Paul {Torrey}, Shy {Genel}, Mark {Vogelsberger}, Ruediger {Pakmor}, Federico {Marinacci}, Rainer {Weinberger}, Luke {Kelley}, Mark {Lovell}, Benedikt {Diemer}, and Lars {Hernquist}.
\newblock {The IllustrisTNG simulations: public data release}.
\newblock \emph{Computational Astrophysics and Cosmology}, 6\penalty0 (1):\penalty0 2, May 2019.
\newblock \doi{10.1186/s40668-019-0028-x}.

\bibitem[Lemos et~al.(2023)Lemos, Coogan, Hezaveh, and Perreault-Levasseur]{lemos2023tarp}
Pablo Lemos, Adam Coogan, Yashar Hezaveh, and Laurence Perreault-Levasseur.
\newblock Sampling-based accuracy testing of posterior estimators for general inference.
\newblock In Andreas Krause, Emma Brunskill, Kyunghyun Cho, Barbara Engelhardt, Sivan Sabato, and Jonathan Scarlett, editors, \emph{Proceedings of the 40th International Conference on Machine Learning}, volume 202 of \emph{Proceedings of Machine Learning Research}, pages 19256--19273. PMLR, 23--29 Jul 2023.
\newblock URL \url{https://proceedings.mlr.press/v202/lemos23a.html}.

\bibitem[{Legin} et~al.(2023){Legin}, {Adam}, {Hezaveh}, and {Perreault-Levasseur}]{Legin2023slic}
Ronan {Legin}, Alexandre {Adam}, Yashar {Hezaveh}, and Laurence {Perreault-Levasseur}.
\newblock {Beyond Gaussian Noise: A Generalized Approach to Likelihood Analysis with Non-Gaussian Noise}.
\newblock \emph{The Astrophysical Journal Letters}, 949\penalty0 (2):\penalty0 L41, June 2023.
\newblock \doi{10.3847/2041-8213/acd645}.

\bibitem[Lu et~al.(2025)Lu, Wu, and Yu]{Lu2025noisy}
Haoye Lu, Qifan Wu, and Yaoliang Yu.
\newblock Stochastic forward–backward deconvolution: Training diffusion models with finite noisy datasets.
\newblock In Aarti Singh, Maryam Fazel, Daniel Hsu, Simon Lacoste-Julien, Felix Berkenkamp, Tegan Maharaj, Kiri Wagstaff, and Jerry Zhu, editors, \emph{Proceedings of the 42nd International Conference on Machine Learning}, volume 267 of \emph{Proceedings of Machine Learning Research}, pages 40741--40768. PMLR, 13--19 Jul 2025.
\newblock URL \url{https://proceedings.mlr.press/v267/lu25n.html}.

\bibitem[Hagemann et~al.(2025)Hagemann, Gruhlke, Stankewitz, Schillings, and Steidl]{Hagemann2025mixed}
Paul Hagemann, Robert Gruhlke, Bernhard Stankewitz, Claudia Schillings, and Gabriele Steidl.
\newblock Provable mixed-noise learning with flow-matching, 2025.
\newblock URL \url{https://arxiv.org/abs/2508.18122}.

\bibitem[O’Riordan and Vegetti(2024)]{ORiordan2024am}
Conor~M O’Riordan and Simona Vegetti.
\newblock Angular complexity in strong lens substructure detection.
\newblock \emph{Monthly Notices of the Royal Astronomical Society}, 528\penalty0 (2):\penalty0 1757--1768, 01 2024.
\newblock ISSN 0035-8711.
\newblock \doi{10.1093/mnras/stae153}.

\bibitem[Bingham et~al.(2018)Bingham, Chen, Jankowiak, Obermeyer, Pradhan, Karaletsos, Singh, Szerlip, Horsfall, and Goodman]{bingham2018pyro}
Eli Bingham, Jonathan~P. Chen, Martin Jankowiak, Fritz Obermeyer, Neeraj Pradhan, Theofanis Karaletsos, Rohit Singh, Paul Szerlip, Paul Horsfall, and Noah~D. Goodman.
\newblock {Pyro: Deep Universal Probabilistic Programming}.
\newblock \emph{Journal of Machine Learning Research}, 2018.
\newblock URL \url{http://jmlr.org/papers/v20/18-403.html}.

\bibitem[Ronneberger et~al.(2015)Ronneberger, Fischer, and Brox]{Ronneberger2015unet}
Olaf Ronneberger, Philipp Fischer, and Thomas Brox.
\newblock U-net: Convolutional networks for biomedical image segmentation.
\newblock In Nassir Navab, Joachim Hornegger, William~M. Wells, and Alejandro~F. Frangi, editors, \emph{Medical Image Computing and Computer-Assisted Intervention -- MICCAI 2015}, pages 234--241, Cham, 2015. Springer International Publishing.
\newblock ISBN 978-3-319-24574-4.
\newblock \doi{10.1007/978-3-319-24574-4_28}.

\bibitem[Song and Ermon(2020)]{Song2020sigma}
Yang Song and Stefano Ermon.
\newblock Improved techniques for training score-based generative models.
\newblock In H.~Larochelle, M.~Ranzato, R.~Hadsell, M.F. Balcan, and H.~Lin, editors, \emph{Advances in Neural Information Processing Systems}, volume~33, pages 12438--12448. Curran Associates, Inc., 2020.
\newblock URL \url{https://proceedings.neurips.cc/paper_files/paper/2020/file/92c3b916311a5517d9290576e3ea37ad-Paper.pdf}.

\bibitem[Lemos et~al.(2025)Lemos, Sharief, Malkin, Salhi, Stone, Perreault-Levasseur, and Hezaveh]{lemos2024}
Pablo Lemos, Sammy~Nasser Sharief, Nikolay Malkin, Salma Salhi, Connor Stone, Laurence Perreault-Levasseur, and Yashar Hezaveh.
\newblock {PQMass}: {Probabilistic} {Assessment} of the {Quality} of {Generative} {Models} using {Probability} {Mass} {Estimation}.
\newblock In \emph{The Thirteenth International Conference on Learning Representations}, 2025.
\newblock URL \url{https://openreview.net/forum?id=n7qGCmluZr}.

\bibitem[Barco et~al.(2025)Barco, Adam, Stone, Hezaveh, and Perreault-Levasseur]{Barco2025}
Gabriel~Missael Barco, Alexandre Adam, Connor Stone, Yashar Hezaveh, and Laurence Perreault-Levasseur.
\newblock Tackling the problem of distributional shifts: Correcting misspecified, high-dimensional data-driven priors for inverse problems.
\newblock \emph{The Astrophysical Journal}, 980\penalty0 (1):\penalty0 108, feb 2025.
\newblock \doi{10.3847/1538-4357/ad9b92}.

\end{thebibliography}

\newpage
\appendix

\section{Strong gravitational-lensing simulator}
\label{app:slsim}

We use \texttt{Caustics} \citep{stone2024caustic} to simulate strong gravitational lensing because it is fully differentiable, allowing us to compute the gradients with respect to both the source and lens parameters required for inference.  
\texttt{Caustics} supports pixelated and parametric descriptions of the source and the lens, and it provides several predefined parametric mass models. In our experiments, the images $\mathbf{x}$, $\mathbf{y}$, and $\boldsymbol{\eta}$ are elements of $\mathbb{R}^{64\times64}$.  Consequently, the Jacobian of the forward model is  
\begin{equation}
    A_{\boldsymbol{\ell}}
    = \nabla_{\mathbf{x}}f(\mathbf{x},\boldsymbol{\ell})
    \in \mathbb{R}^{4096\times4096}.
\end{equation}
The lens mass is modeled as an Elliptical Power-Law (EPL) profile with external shear and $m=3$ multipole.

Our forward model also includes a Gaussian point-spread function (PSF) with $\mathrm{FWHM}=0.375^{\prime\prime}$.  In principle, the simulator first applies the lensing transformation and then convolves the result with the PSF.  Because convolution with a fixed kernel is a linear operation (represented by a circulant matrix) and the PSF parameters are held constant, we denote the combined lensing-plus-PSF operator simply by $A_{\boldsymbol{\ell}}$.

The lens and source redshifts are assumed to be known and fixed at $z_\ell=0.5$ and $z_s=1.0$, respectively.  
The source plane has a field of view (FOV) of $6.24^{\prime\prime}$, while the observation plane spans $12^{\prime\prime}$.  
The $6.24^{\prime\prime}$ source FOV corresponds to $\simeq50\,\mathrm{kpc}$ at $z=1$, matching the window used when training the galaxy prior, whereas the $12^{\prime\prime}$ observation window is wide enough to encompass all simulated lenses and is consistent with the typical angular extent of real strong-lensing systems. Finally, we add Gaussian additive noise to the simulations, with $\boldsymbol{\eta} \sim \mathcal{N}(0, \sigma_\eta^2\mathbb{I})$ and $\sigma_\eta = 0.35$.

\section{Uniform prior ranges}
\label{app:priors}

\Tab{prior_ranges} lists the uniform priors used for both the simulations and the inference runs.  
The intervals are chosen to encompass the bulk of galaxy–scale strong lenses reported in the literature.  
All distance‐related quantities are expressed in arcseconds, and we follow the \texttt{Caustics} convention for angles (measured counter-clockwise from the positive $x$-axis) and for all other parameters.

The $m{=}3$ multipole amplitude is kept small, consistent with previously reported values \citep[e.g.][]{ORiordan2024am}.  
Although $a_{m}$ is weakly constrained in our experiments, and a wider inference prior would explore the parameter space more thoroughly, we retain the same interval for simulation and inference so that a future coverage test remains well defined.
\begin{table}[htb]
    \centering
    \begin{tabular}{|c|c|c|c|}
        \hline
        Plane & Component & Parameter & Uniform prior \\
        \hline\hline
        \multirow{10}{*}{Lens} 
            & \multirow{2}{*}{Lens center} 
                & $x_\ell$ & $[-0.25,\,0.25]$ \\
            \cline{3-4}
            & & $y_\ell$ & $[-0.25,\,0.25]$ \\
            \cline{2-4}
            & \multirow{4}{*}{EPL profile} 
                & $q$                & $[0.70,\,1.00]$ \\
            \cline{3-4}
            & & $\phi$              & $[0,\,\pi]$ \\
            \cline{3-4}
            & & $\theta_\mathrm{E}$ & $[1,\,3]$ \\
            \cline{3-4}
            & & $\tau$              & $[0.75,\,1.25]$ \\
            \cline{2-4}
            & \multirow{2}{*}{External shear} 
                & $\gamma_1$ & $[-0.25,\,0.25]$ \\
            \cline{3-4}
            & & $\gamma_2$ & $[-0.25,\,0.25]$ \\
            \cline{2-4}
            & \multirow{2}{*}{Multipole $m{=}3$} 
                & $a_{m}$   & $[0.00,\,0.015]$ \\
            \cline{3-4}
            & & $\phi_{m}$ & $[0,\,2\pi]$ \\
        \hline
        \multirow{7}{*}{Source} 
            & \multirow{2}{*}{Source center} 
                & $x_{s}$ & $[-0.35,\,0.35]$ \\
            \cline{3-4}
            & & $y_{s}$ & $[-0.35,\,0.35]$ \\
            \cline{2-4}
            & \multirow{5}{*}{Sérsic light$^{\dagger}$}
                & $q_{s}$   & $[0.05,\,1.00]$ \\
            \cline{3-4}
            & & $\phi_{s}$ & $[0,\,\pi]$ \\
            \cline{3-4}
            & & $n_{s}$    & $[0.3,\,10]$ \\
            \cline{3-4}
            & & $R_{s}$    & $[0.1,\,3]$ \\
            \cline{3-4}
            & & $I_{s}$    & $[0.6,\,100]$ \\
        \hline
    \end{tabular}
    \vspace{0.2cm}
    \caption{Uniform priors for all parameters used in the simulations.  
    $^{\dagger}$The Sérsic source parameters are used only to initialize $\boldsymbol{\ell}_0$.}
    \label{tab:prior_ranges}
\end{table}

\section{Marginal lens posterior}
\label{app:marginal_posterior}

\begin{figure*}[!htb]
    \centering
    \includegraphics[width=3.25in]{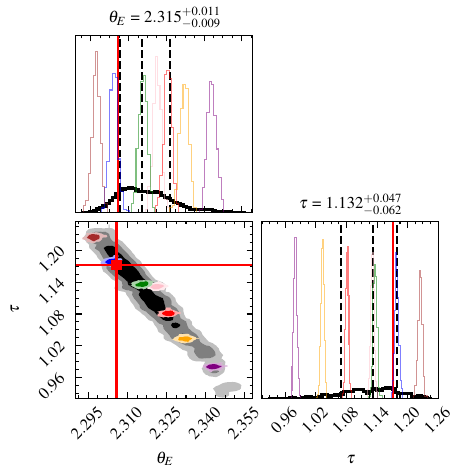}
    \caption{Marginal lens posterior $p(\boldsymbol{\ell}\mid\mathbf{y})$ (black contours and histograms) compared with six conditional posteriors $p(\boldsymbol{\ell}\mid\mathbf{y},\mathbf{x})$ (colored curves), where each $\mathbf{x}$ is drawn from the joint sampler.  Only the parameters $\theta_{\mathrm{E}}$ and $\tau$ are shown for clarity.}
    \label{fig:marginal_vs_conditional_lens}
\end{figure*}

To estimate the marginal lens posterior, we write
\begin{align}
    p(\boldsymbol{\ell} \mid \mathbf{y}) 
    & = \int_{\mathbf{x}_0} p(\boldsymbol{\ell}, \mathbf{x}_0 \mid \mathbf{y}) \,d\mathbf{x}_0 = \int_{\mathbf{x}_0} p(\boldsymbol{\ell} \mid \mathbf{x}_0, \mathbf{y}) \,p(\mathbf{x}_0 \mid \mathbf{y})\,d\mathbf{x}_0 \nonumber \\
    & =  \int_{\mathbf{x}_0} \int_{\bar{\boldsymbol{\ell}}} p(\boldsymbol{\ell} \mid \mathbf{x}_0, \mathbf{y}) \,p(\mathbf{x}_0, \bar{\boldsymbol{\ell}} \mid \mathbf{y})\,d\bar{\boldsymbol{\ell}}\,d\mathbf{x}_0 \\ & = \mathbb{E}_{(\mathbf{x}_0, \bar{\boldsymbol{\ell}}) \sim p(\mathbf{x}_0, \bar{\boldsymbol{\ell}} \mid \mathbf{y})}\bigl[p(\boldsymbol{\ell} \mid \mathbf{x}_0, \mathbf{y})\bigr] \\
    & \approx \frac{1}{n}\sum_{i = 1}^n p\bigl(\boldsymbol{\ell} \mid \mathbf{x}_0^{(i)}, \mathbf{y}\bigr),
\end{align}
where $\mathbf{x}_0^{(i)}\sim p(\mathbf{x}_0,\bar{\boldsymbol{\ell}}\mid\mathbf{y})$ are samples from the joint sampler.  
We sample the conditional density with the NUTS sampler \citep{Hoffman2014nuts} as implemented in \texttt{Pyro}~\citep{bingham2018pyro}.

This strategy is more efficient than drawing a very large number of joint samples: we obtain a dense estimate of the marginal by running short conditional chains for several fixed sources.  In addition, the conditionals $p(\mathbf{x}\mid\mathbf{y},\boldsymbol{\ell})$ and $p(\boldsymbol{\ell}\mid\mathbf{y},\mathbf{x})$ are better understood than the approximate joint sampler, which still awaits a formal coverage study.

\Fig{marginal_vs_conditional_lens} shows the resulting marginal posterior (black) together with six conditional posteriors (colors) from different source draws.  Each conditional distribution is much narrower and nested within the marginal, illustrating how source–lens coupling broadens the overall posterior.  Relying on a single point estimate for $\boldsymbol{\ell}$ without fully exploring the joint posterior can therefore lead to biased scientific conclusions. For the marginal lens posteriors displayed in this work, we get $406$ joint samples, and $500$ conditional lens posterior samples per source.

\section{Score-based model: architecture and training}
\label{app:sbm_training}

We train a variance-exploding (VE) score model on the galaxy dataset described in \Sec{experiments}, which contains simulated galaxy images $\mathbf{x}\in\mathbb{R}^{64\times64}$ covering a $50\,\mathrm{kpc}$ window.  
Model definition and training are implemented with the \texttt{score-models}\footnote{\href{https://github.com/AlexandreAdam/score_models}{github.com/AlexandreAdam/score\_models}} package.  
The network $\mathbf{s}_\theta(\mathbf{x},t)$ is a noise-conditional score network \citep{Song2019ve} with a U-Net architecture \citep{Ronneberger2015unet}.

\paragraph{Network hyperparameters.}
Within \texttt{score-models} we adopt \text{"nf": 64}, \text{"ch\_mult": [1, 2, 2, 3]}, and \text{"num\_blocks": 3}, and leave all other settings at their defaults.

\paragraph{Optimization.}
The model is trained with Adam \citep{kingma2017adam} using a learning rate $5\times10^{-5}$, an EMA decay of $0.999$, a batch size of $256$, and $1000$ epochs ($\approx230\,000$ optimization steps).  
Training required 45 h on a single A100\,40GB GPU.

\paragraph{SDE parameters and data normalization.}
We parametrize time as $t\in[0,1]$ with $\sigma_{\min}=0.001$ and $\sigma_{\max}=50$.  
The data are normalized as $(\mathbf{x}-M)/C$ with $M=0.125$ and $C=10.115$, so that most pixel values lie below~1.  
Following \citet{Song2020sigma}, the maximum noise level is set from pair-wise distances; we use the 95th percentile (rather than the maximum) to avoid the undue influence of a few very bright galaxies.

\paragraph{Statistical fit.}
We evaluate the learned prior with \textsc{PQMASS} \citep{lemos2024}, a sample-based $\chi^{2}$ test that divides the data space into $n$ regions and estimates the densities of both sample sets in each region. The regions are defined by randomly selecting $n$ reference points from one sample set and constructing the corresponding Voronoi tessellation. The resulting statistic, $\chi^{2}_{\mathrm{PQM}}$, follows a $\chi^{2}$ distribution with $n-1$ degrees of freedom; we report its value averaged over $m$ random re-tessellations. Using $4000$ prior samples, $n = 100$ regions, $m = 2000$ re-tessellations, and $4000$ training images, we obtain $\chi^2_{\mathrm{PQM}} = 104.39 \pm 13.19$; comparing to $4000$ validation images yields $\chi^2_{\mathrm{PQM}} = 116.2 \pm 14.1$. Both values are close to the expected mean (99) of the $\chi^2_{(99)}$ distribution, indicating that the learned score model faithfully represents the galaxy distribution.

\section{Inference hyperparameters and implementation details}
\label{app:inference_hyper}

\paragraph{Source-only inference (non-blind inversion).}
For source-only inference, we adopt the Convolved Likelihood Approximation because it has been shown to satisfy the TARP coverage test \citep{lemos2023tarp} under appropriate noise levels, solver accuracy, and prior choice \citep{Barco2025}.  
In our experiments, CLA yields residual $\chi^{2}$ values clustered around the ground-truth noise realization (to which we have access).  
We integrate the SDE with 1\,000 steps.

\paragraph{Schedule for $r_t^{2}$.}
For both CLA and $\Pi$GDM we set
\begin{equation}
    r_t^{2}= \sigma(t)^{2}\!\bigl(C^{2}t^{4}+1\bigr),
    \label{equ:r_t}
\end{equation}
instead of the original choices $r_t^{2}=\sigma(t)^{2}$ (CLA) or $r_t^{2}=\sigma(t)^{2}/\bigl[\sigma(t)^{2}+1\bigr]$ ($\Pi$GDM).  
The original schedules often lead to unstable diffusion, especially for bright, high-S/N galaxies, whereas the schedule in \Equ{r_t} performs robustly in all our tests.  
The boundary condition at $t\!\to\!0$ remains exact, so the score approximation is unchanged at the data end of the SDE.  
A formal ablation study of $r_t^{2}$, ideally using a coverage metric such as TARP, is left for future work.

\begin{figure*}[!htb]
    \centering
    \includegraphics[width=5.5in]{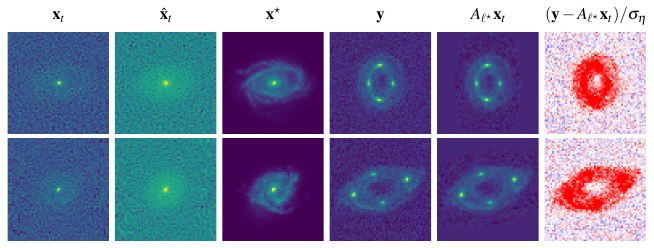}
    \caption{Source-only inference variables with CLA at $t=0.5$.}
    \label{fig:cla_example}
\end{figure*}

\begin{figure*}[!htb]
    \centering
    \includegraphics[width=5.5in]{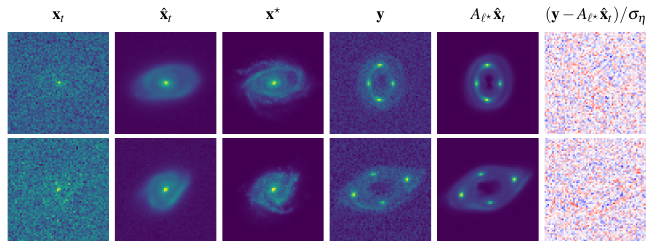}
    \caption{Source-only inference variables with $\Pi$GDM at $t=0.5$.}
    \label{fig:pgdm_example}
\end{figure*}

\paragraph{Joint inference.}
For blind inversion, we prefer $\Pi$GDM, because the modifications CLA makes to $\mathbf{x}_{t}$ degrade the Tweedie estimate $\hat{\mathbf{x}}_{t}$ required for the lens update.  
\Fig{cla_example} and \Fig{pgdm_example} compare the two methods at $t=0.5$: the $\Pi$GDM estimate of $\hat{\mathbf{x}}_{t}$ resembles a realistic galaxy and lies close to the ground truth, whereas the CLA estimate is noticeably degraded.  
With CLA, the joint sampler often diverges; with $\Pi$GDM, it converges reliably.

\begin{figure*}[!htb]
    \centering
    \includegraphics[width=5.5in]{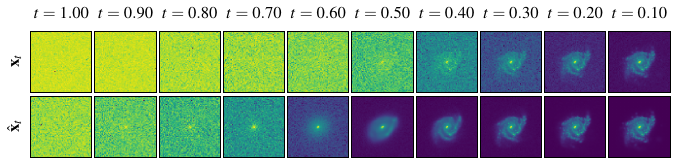}
    \caption{Evolution of the Tweedie posterior mean $\hat{\mathbf{x}}_{t}$ during $\Pi$GDM source-only inference.}
    \label{fig:tweedies}
\end{figure*}

We run 400 reverse-diffusion steps and, at each, perform 500 ULA updates of the lens parameters with a step size $10^{-7}$ to limit discretization bias.  
The lens is updated only for $t\in[0.7,\,0.2]$: at high $t$ the Tweedie estimate is inaccurate, and for $t<0.2$ the source has essentially converged \citep[see also][]{murata2023gibbs}.  
\Fig{tweedies} illustrates the evolution of $\mathbf{x}_{t}$ and $\hat{\mathbf{x}}_{t}$ throughout the diffusion process for an experiment doing source-only inference.

\paragraph{Lens initialization and sampling.}
We obtain $\boldsymbol{\ell}_{0}$ by minimizing the negative log-likelihood under a Sérsic source model.  
Adam \citep{kingma2017adam} is run from 1\,250 random starts drawn from the priors in \App{priors}, using a learning rate of $0.25$ and 8\,000 optimization steps. For conditional lens sampling, we use NUTS with an initial step size $10^{-3}$, 750 warm-up steps, and a maximum tree depth of 9.

\paragraph{Computation time.}
All experiments were performed on a single A100 GPU (40 GB).  
Approximate wall-times are:
\begin{itemize}
    \item Lens initialization: $\sim$30 min;
    \item Joint sampling of seven $(\mathbf{x}_{i},\boldsymbol{\ell}_{i})$ pairs (parallel): $\sim$40 min;
    \item Source-only sampling (seven sources): $\sim$5 min;
    \item Conditional lens sampling with NUTS for seven lens conditional posteriors: $\sim$1 h (varies with posterior complexity).
\end{itemize}

\section{Complete lens posteriors and additional experiments}
\label{app:extra_experiments}

\begin{figure*}[!htb]
    \centering
    \includegraphics[width=5.5in]{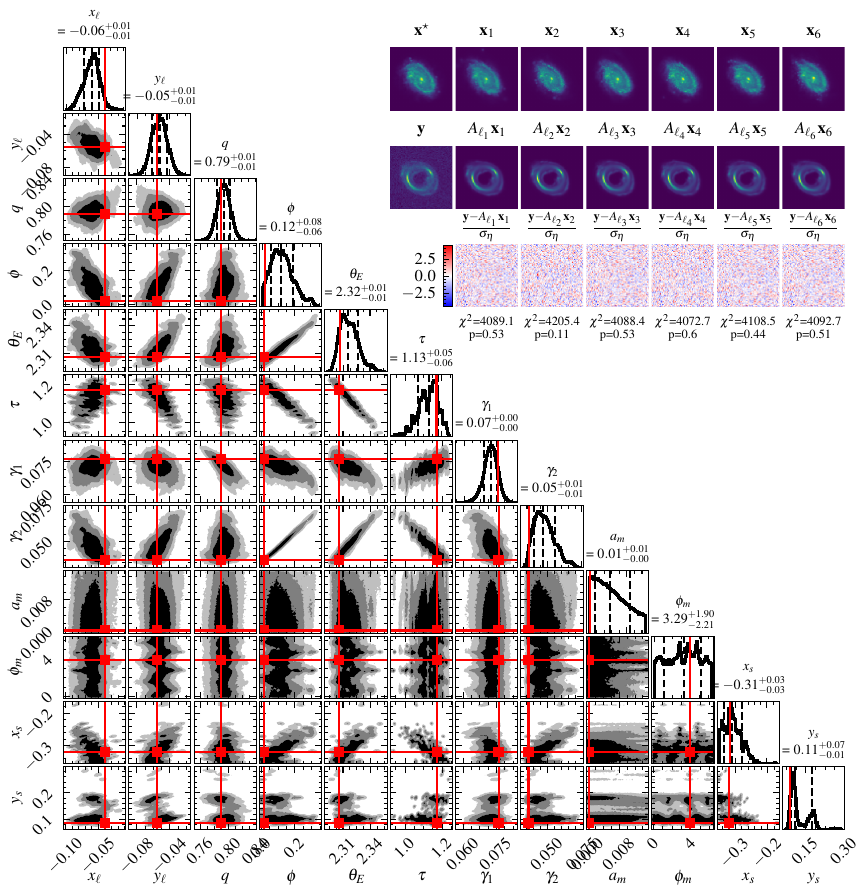}
    \caption{Full marginal posterior for the 12 lens parameters in the main experiment, along with several source–lens reconstructions and residuals.}
    \label{fig:exp0}
\end{figure*}

\begin{figure*}[!htb]
    \centering
    \includegraphics[width=5.5in]{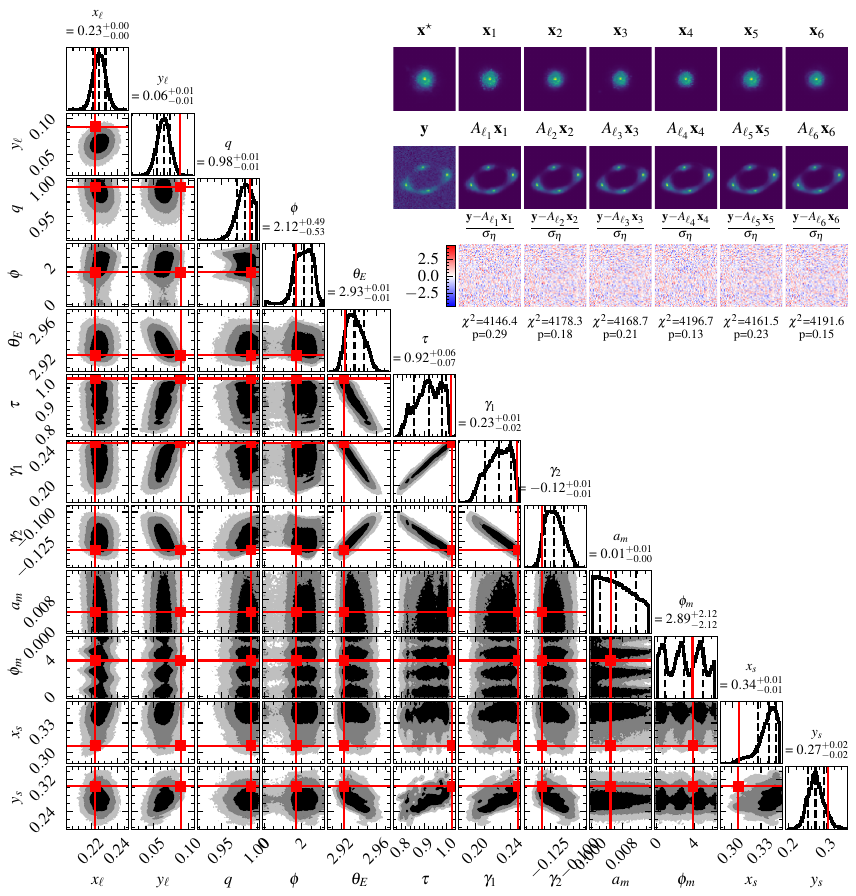}
    \caption{Experiment~2: marginal lens posterior and joint samples for a second simulated system.}
    \label{fig:exp1}
\end{figure*}

\begin{figure*}[!htb]
    \centering
    \includegraphics[width=5.5in]{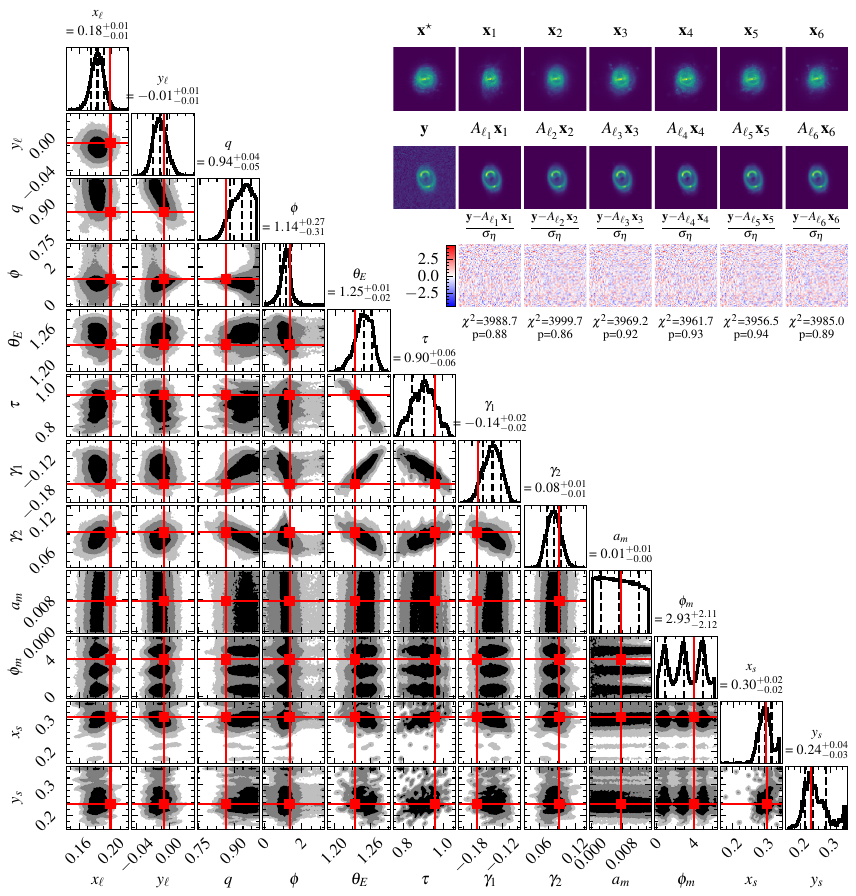}
    \caption{Experiment~3: marginal lens posterior and joint samples for a second simulated system.}
    \label{fig:exp2}
\end{figure*}

\begin{figure*}[!htb]
    \centering
    \includegraphics[width=5.5in]{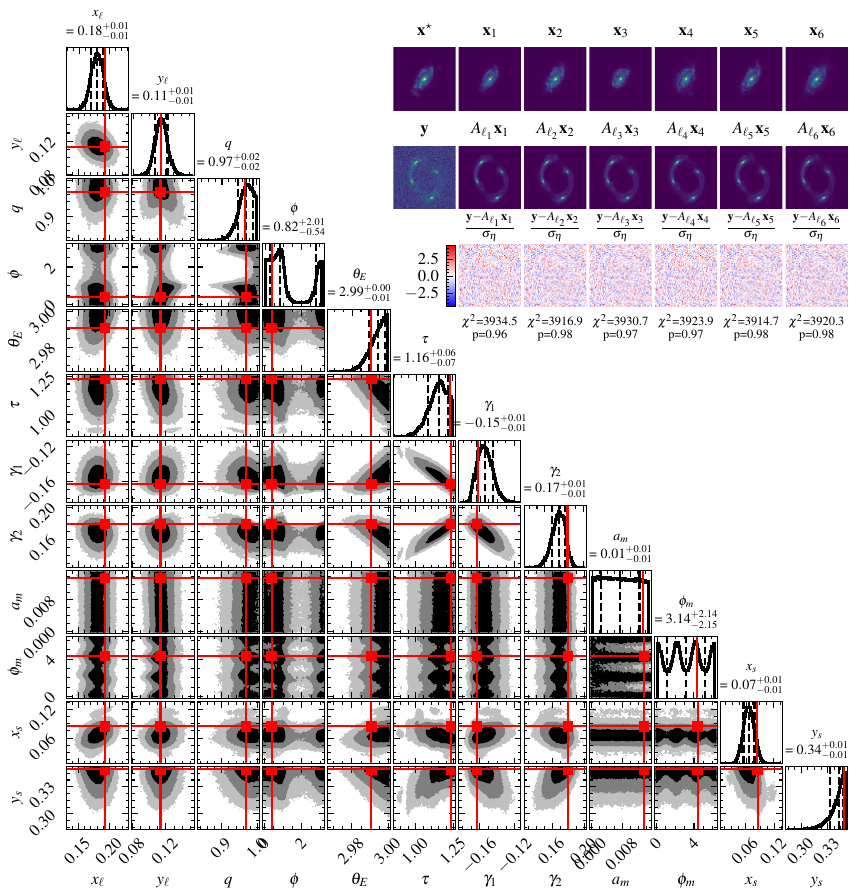}
    \caption{Experiment~4: marginal lens posterior and joint samples for a third simulated system.}
    \label{fig:exp3}
\end{figure*}

\Fig{exp0} presents the full marginal posterior for all 12 macro parameters in the main experiment, together with additional source samples, their lens reconstructions, and residual maps.  
To demonstrate robustness, we include three further simulated systems in \Fig{exp1}, \Fig{exp2}, and \Fig{exp3}.  
Each figure follows the same layout: the corner plot of the marginal lens posterior, ground-truth lens values (red markers), and several joint posterior draws with their corresponding residuals.

\end{document}